\documentclass[sigconf]{acmart}

\usepackage{balance}
\usepackage{multirow}
\usepackage{makecell}

\AtBeginDocument{%
  }

\copyrightyear{2024} 
\acmYear{2024} 
\setcopyright{acmlicensed}
\acmConference[WWW '24 Companion]{Companion Proceedings of the ACM Web Conference 2024}{May 13--17, 2024}{Singapore, Singapore}
\acmBooktitle{Companion Proceedings of the ACM Web Conference 2024 (WWW '24 Companion), May 13--17, 2024, Singapore, Singapore}
\acmDOI{10.1145/3589335.3651500}
\acmISBN{979-8-4007-0172-6/24/05}

\begin{document}

\title{Event GDR: Event-Centric Generative Document Retrieval}






\author{Yong Guan}
\affiliation{%
  \institution{Tsinghua University}
  \state{Beijing}
  \country{China}
  }
\email{gy2022@mail.tsinghua.edu.cn}

\author{Dingxiao Liu}
\author{Jinchen Ma}
\affiliation{%
  \institution{Zhipu.AI}
  \state{Beijing}
  \country{China}
  }

\author{Hao Peng}
\author{Xiaozhi Wang}
\affiliation{%
  \institution{Tsinghua University}
  \state{Beijing}
  \country{China}
  }

\author{Lei Hou}
\authornote{Corresponding Author}
\affiliation{%
  \institution{Tsinghua University}
  \state{Beijing}
  \country{China}
  }
\email{houlei@tsinghua.edu.cn}

\author{Ru Li}
\affiliation{%
  \institution{Shanxi University}
  \state{Shanxi}
  \country{China}
  }
\email{liru@sxu.edu.cn}


\renewcommand{\shortauthors}{Yong Guan et al.}

\begin{abstract}
Generative document retrieval, an emerging paradigm in information retrieval, learns to build connections between documents and identifiers within a single model, garnering significant attention. However, there are still two challenges: (1) \textit{neglecting inner-content correlation during document representation}; (2) \textit{lacking explicit semantic structure during identifier construction}. Nonetheless, events have enriched relations and well-defined taxonomy, which could facilitate addressing the above two challenges. Inspired by this, we propose Event GDR, an event-centric generative document retrieval model, integrating event knowledge into this task. Specifically, we utilize an exchange-then-reflection method based on multi-agents for event knowledge extraction. For document representation, we employ events and relations to model the document to guarantee the comprehensiveness and inner-content correlation. For identifier construction, we map the events to well-defined event taxonomy to construct the identifiers with explicit semantic structure. 
Our method achieves significant improvement over the baselines on two datasets, and also hopes to provide insights for future research.


\end{abstract}

%
%


%


\maketitle

\section{Introduction}
\label{sec:intro}

\begin{figure}[t]
    \centering
    \includegraphics[width=0.8\linewidth]{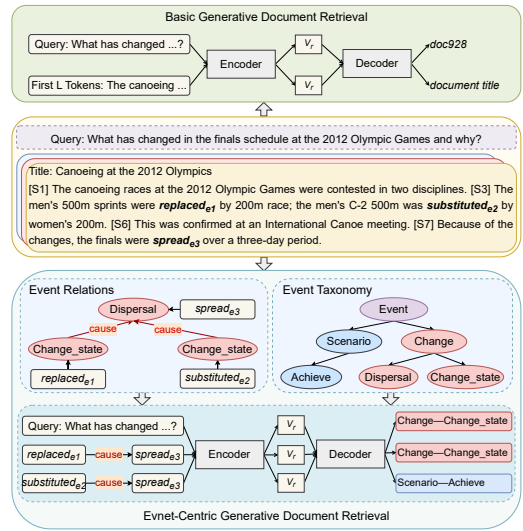}
    \caption{Basic GDR (top) and Event-Centric GDR (bottom).}
    \label{fig:intro_example}
\end{figure}

Document retrieval (DR) aims to identify and provide a list of relevant documents from an extensive document collection or repository in response to a user's query. 
It is a fundamental subfield of information retrieval, and plays a pivotal role in many real-world applications. 
Especially with the advent of large language models (LLMs), DR becomes even more important, as it can facilitate LLMs to solve various problems such as hallucinations and reasoning. 
Existing document retrieval systems mainly follow an index-then-retrieval pipeline. However, this two-stage pipeline is difficult to jointly optimize all heterogeneous modules.

Recently, an alternative paradigm has been explored, namely generative document retrieval (GDR)~\cite{tay2022transformer,pradeep2023does},  which investigates generating relevant document identifiers for a given query using a single model. 
However, there are still two critical challenges. First, 
\textbf{neglecting inner-content correlation during document representation}. 
Document representation, a main step in GDR, aims to store document information into parameters~\cite{zhuang2022bridging}. 
Existing methods typically use key content such as distinctive substring~\cite{tay2022transformer} for document modeling, which lacks contextualized information. 
Although~\citet{yubao_2023_kdd} try to model independent chunks of document, they neglect the inner-content correlation. Intuitively, considering inner-content correlation will be better for document modeling. 
Second,  
\textbf{lacking explicit semantic structure during identifier construction}. 
The quality of identifier construction is an important factor that affects model performance~\cite{li-etal-2023-multiview}. 
Several types of identifiers have been explored, such as unique integer~\cite{tay2022transformer} and titles~\cite{li-etal-2023-multiview} (Figure \ref{fig:intro_example}), which ignore structural information. While~\citet{wang2022neural} utilize numeric strings by cluster, a gap between the numeric and content exists, impeding understanding and indexing.  

Interestingly, events have enriched relations and well-defined taxonomy, which could facilitate addressing the above two challenges. 
\textit{For document representation}, enriched event relations are able to help model inner-content correlation. A document typically contains multiple events that can convey its main content and enhance the comprehensiveness of modeling. Besides, events are not isolated, and they have enriched relations, such as causal and sub-event. Different contents of the document can be connected through event relations. For example, query in Figure \ref{fig:intro_example} can be easily answered while knowing the causal relations among events 1, 2, and 3. 
\textit{For identifier construction}, well-defined taxonomy can provide useful hierarchical structure information for building identifiers. 
Each event type is a node in taxonomy and has hierarchical relations with other nodes, such as \texttt{Dispersal} and \texttt{Change\_tool} are the subclasses of \texttt{Change}. 
In addition, event taxonomy is a structured abstraction of event instances, closely linked to them, thereby deepening the connection between identifiers and document.

In this paper, we propose Event GDR, an event-centric generative document retrieval model, which has the best of both worlds, by building on enriched event relations and well-defined taxonomy for generative document retrieval. 
Specifically, inspired by the fundamental characteristics of human problem-solving, i.e., exchange, reflection, we first utilize an exchange-then-reflection method for event knowledge extraction, in which multiple language model instances (multi-agents) communicate with each other to obtain external insights~\cite{du2023improving}, and one of them iteratively reflects~\cite{noah_2023_reflexion} based on the insights to reach the final responses. Then, we utilize events and relations to model the document, which could guarantee the comprehensiveness and inner-content correlation. Finally, we map the events to well-defined  taxonomy to construct the  identifiers with explicit semantic structure. The contributions are as follows:

\begin{itemize}
    \item We propose Event GDR, an event-centric generative document retrieval model that attempts to leverage the enriched event relations and well-defined taxonomy to address the two challenges on document representation and identifier construction for generative document retrieval.
    \item We employ an exchange-then-reflection method, where multiple agents facilitate the event knowledge extraction through communication and reflection.
    \item Experiments on Natural Questions and DuReader datasets show that our method equipped with event relations and taxonomy consistently improves the performance.
\end{itemize}

\section{Preliminaries}
\label{sec_preliminary}

This section aims to describe the basic modes of the GDR, which performs index and retrieval within a single model.

\textbf{Indexing Phase} aims to store the documents into the model parameters and build the relations between documents and identifiers. 
Given a document $d_{i} \in D$ and corresponding document identifier $z$. Assume $f$ is an autoregressive language model, e.g., T5~\cite{raffel_t5_2020}, which takes the document content as input and outputs identifiers. 
\begin{equation}
    \mathcal{L}_{index} = \sum_{d_{i} \in D} log P(z|f_{\theta}(d_{i}))
\end{equation}\label{eq_ori_index}
\textbf{Retrieval Phase} aims to return relevant document identifier given a query $q$. The probability of identifier $z$ can be computed as: 
\begin{equation}
    P(z|q, \theta) = \prod_{t=1}P(z_{i}|f_{\theta}(q, z_{<t}))
\end{equation}\label{eq_ori_retrieval}

\begin{figure}
    \centering
    \includegraphics[width = \linewidth]{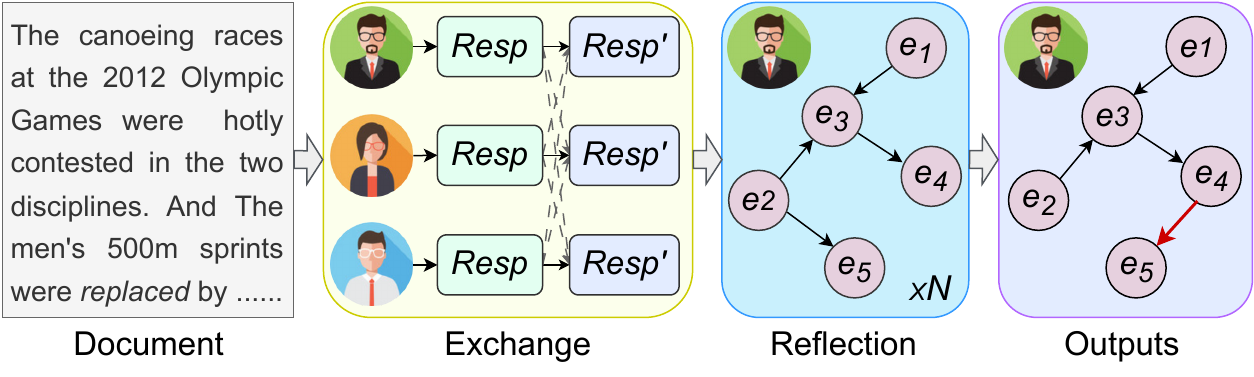}
    \caption{Event/Relation Extraction Process.}
    \label{fig:event_extraction}
\end{figure}

\section{Methodology}


\subsection{Event and Relation Extraction}
\label{sec_event_extraction}

This section aims to extract the events and relations in document to convey the main content. Inspired by the  characteristics of human problem-solving, i.e., exchange, reflection, we propose to extract events and relations subject to an exchange-then-reflection (ExR) procedure based on multiple agents. Given a document, multi-agents offer external insights through communication~\cite{du2023improving}, and one of them iteratively reflects~\cite{noah_2023_reflexion} based on insights to find the final answers. As shown in Figure \ref{fig:event_extraction}, ExR comprises two steps.

\textbf{Exchange} aims to obtain external insights by exchanging information with each other. Specifically, we initially prompt each agent to autonomously extract relevant events or relations from the document. Subsequently, the insights gathered by each agent are shared across the collective network, fostering an environment of collaborative intelligence. Finally, every agent thoughtfully refines their initial response, incorporating the diverse perspectives and feedback obtained from other agents.

\textbf{Reflection} aims to reflect based on the insights provided by different agents to give the final responses. After obtaining external insights from each agent, one of the agents continuously refines responses based on these insights until the ``done'' function is invoked or the iteration cap is reached.

Event extraction and event relation extraction are two independent phases. For event extraction, each event is a tuple <trigger words, mention>, where mention is the sentence describing the event, and trigger word is the word evoking the event. The model restates and simplifies sentences of event mention to remove redundant elements like irrelevant examples. For relation extraction, both the document and the extracted events serve as prior information. The model extracts the relations, focusing on causal relation between each pair of events.

\subsection{Document Representation}
\label{sec_document_reps}

Document representation aims to utilize the events and relations to represent the document. 

\textbf{Event as Representation (EReps)} focuses on representing the document through events. We use the mention of each event, as extracted from section \ref{sec_event_extraction}, to separately represent the document. From a comprehensiveness perspective, we consider all event mentions, and associate them with corresponding identifiers.

\textbf{Event Relation as Representation (ERReps)} maps events with relations instead of individual event to identifiers. For each event pair with relations, we use event mentions that are ordered by the relation direction to represent the document. Afterward, we could obtain a set of event and relation representations (denoted as $E_{i} = \{e_{i}\}$ and $R_{i} = \{r_{i}\}$) for individual document $d_{i}$. The advantage of utilizing event and relation is considering both the comprehensiveness and the inner-content correlation.

\subsection{Identifier Construction}
\label{sec_identifier_cons}

This section aims to build identifiers with explicit semantic structures and explores two ways for representing identifiers.

\textbf{Event as Identifiers (EIds)} \ aims to adopt events with a clustering algorithm to build identifiers. We utilize events for clustering, each document is assigned a numerical identifier, indicating that events or relations within the document share the same identifier. Unlike previous work that uses the first $L$ tokens \cite{wang2022neural} (denoted as \textbf{TIds}), exploiting events can cover the main content of a document.

\textbf{Event Taxonomy as Identifiers (ETIds)} \ aims to map the events to taxonomy and use the hierarchical structure to build identifiers with explicit semantic structures. The semantic identifier for the content is formed by concatenating the event types along the path from the root event to its corresponding leaf event type. 
To achieve this, we need to map the event to corresponding event type in taxonomy. The mapping for $e_{i}$ can be computed as: 
\begin{equation}
\mathit{ET}(e_{i}, \mathcal{T}) = \arg\max ( \sum_{v_k \in V_{i}} \rho(v_{k}, \mathcal{T}) )
\end{equation}
where $v_{k}$ is the $k$-th feature for $V_{i}$, $\mathcal{T}$ is the taxonomy, $V_{i}$ is the feature set of $e_{i}$, $\rho(\cdot)$ can be an arbitrary score function for $v_{k}$ and $\mathcal{T}$, such as deep neural network, a simple linear projection.

\subsection{Model Training}
\label{sec_model_training}

The module aims to introduce the training strategy. We replace the original document $d_{i}$ that need to be indexed in Equation \ref{eq_ori_index} with their corresponding events $e_{i}$ and event relations $r_{i}$. We utilize multi-task settings to learn indexing and retrieval simultaneously. 
\begin{equation}
\begin{split}
    \mathcal{L} =& \sum_{d_{i} \in D} \sum_{e_{i} \in E_{i}}  log P(z|f_{\theta}(e_{i})) + \sum_{q_{i} \in Q} log P(z|f_{\theta}(q_{i})) + \\ & \sum_{d_{i} \in D} \sum_{r_{i} \in R_{i}}  log P(z|f_{\theta}(r_{i}))
\end{split}
\end{equation}

\begin{table}[t]
    \small
    \centering
        \caption{Overall Retrieval Performance. Event GDR represents the best-performing variant ERReps+ETIds.}
    \begin{tabular}{l|cccccc}
    \toprule
    \multirow{2}*{Methods}& \multicolumn{3}{c}{NQ\_10K}& \multicolumn{3}{c}{DuReader\_50K}\\
    &@1&@10&@20&@1&@10&@20\\
    \midrule
    BM25&40.16&52.79&61.31&44.39&57.25&67.41\\
    DPR~\cite{karpukhin-etal-2020-dense}&62.09&70.25&68.71&60.14&73.92&76.30\\
    GAR~\cite{mao-etal-2021-generation}&58.26&69.23&67.55&59.56&75.48&78.31\\
    \midrule
    DSI~\cite{tay2022transformer}&17.56&21.30&22.61&1.12&4.35&5.41\\
    DSI-QG~\cite{zhuang2022bridging}&60.16&68.04&68.78&60.11&73.22&76.12\\
    SE-DSI~\cite{yubao_2023_kdd}&63.82&70.56&70.81&59.86&73.77&76.26\\
    NCI~\cite{wang2022neural}&66.01&73.41&74.63&56.20&72.82&76.67\\
    DSI-QG-SEM~\cite{chen-etal-2023-understanding}&70.48&76.26&76.99&61.71&76.06&78.47\\
    \midrule
    \textit{Ours}\\
    EReps + TIds &65.86&75.44&76.09&60.06&76.37&79.07\\
    EReps + EIds &65.69&75.36&76.66&60.26&77.12&80.18\\
    EReps + ETIds &70.22&80.08&81.12&63.01&\textbf{80.18}&83.38\\
    ERReps + TIds &66.34&75.85&76.50&61.11&76.42&79.73\\
    ERReps + EIds &66.91&75.52&77.07&61.31&77.47&80.23\\ 
    Event GDR  &\textbf{73.25}&\textbf{80.65}&\textbf{82.03}&\textbf{63.96}&80.13&\textbf{83.49}\\
    \bottomrule
    \end{tabular}
    \label{tab:main_results}
\end{table}

\section{Experiments}

\subsection{Evaluation Setup}

\noindent \textbf{Datasets and Metrics.} \ \ We conduct experiments on two widely used datasets with different languages, namely Natural Question, for English, and DuReader\_retrieval, for Chinese. Natural Question (NQ) contains query-document pairs. Similar to existing work~\cite{tay2022transformer,yubao_2023_kdd}, we randomly sample $\approx10K$ query-document pairs to form NQ\_10K. DuReader\_retrieval, a passage retrieval dataset, aims to find relevant passages for the question, which has multiple candidates. 
We combine these candidates into a single document and randomly sample $\approx50K$ query-document pairs to form DuReader\_50K. Following the settings of GDR, we use Hits@\{1,10,20\} as evaluation metrics for both datasets.

\noindent \textbf{Implementation Details.} \ \ We use the pretrained T5~\cite{raffel_t5_2020} and multi-lingual T5~\cite{xue-etal-2021-mt5} models to initialize the model parameters for NQ\_10K and DuReader\_50K respectively. \textit{For event knowledge extraction}, we employ ChatGPT and GLM-130B during exchange phase, and select ChatGPT during the reflection phase. 
\textit{For taxonomy selection}, we select MAVEN's taxonomy as it is more generalized and shared data source with NQ\_10K, both originating from Wikipedia. We translate taxonomy into Chinese for Dureader\_50K to also test the model's generalizability. 
\textit{Regarding the event-to-taxonomy mapping}, for NQ\_10K, following existing work~\cite{guan-etal-2023-trigger}, we train a neural model on MAVEN for mapping. For Dureader\_50K, take event mentions, definitions, event types, as features and use pre-trained models, such as BERT, to obtain feature representations, selecting the most relevant type for the event.

\subsection{Main Results}
The overall results are shown in Table \ref{tab:main_results}, \textbf{among the different types of retrieval methods, our Event GDR equipped with event knowledge achieves the best performance on both datasets. This validates the effectiveness of enhancing GDR through event knowledge extracted by LLMs, also providing an insight for subsequent research.} By jointly analyzing the performance, we would like to answer the challenges for GDR.

\textbf{Do events and relations benefit document representation? } \ \textit{\textbf{Yes.  We observe that utilizing events and relations more effectively represents documents, significantly enhancing performance with a 7.7\% improvement.}} First, \textit{EReps+TIds} outperforms \textit{DSI-QG}, primarily due to \textit{EReps+TIds} representing document with events, while \textit{DSI-QG} utilizes pseudo-queries. Second, integrating event relations further boosts model performance, achieving better performance on DuReader\_50K compared to all baselines.

\textbf{Does event taxonomy benefit identifier construction? } \ \textit{\textbf{Yes. Our method, \textit{Event GDR}, achieves optimal performance on both datasets by integrating event taxonomy, effectively enhancing the association between identifiers and document. }} The table shows that integrating taxonomy into various methods enhances model performance, and the variant (\textit{ERReps+ETIds}) combining event relations and taxonomy achieves the best results.

\begin{figure}[t]
\centering
\begin{tabular}{cc}
\begin{minipage}[t]{0.45\linewidth}
   \includegraphics[width = 1\linewidth]{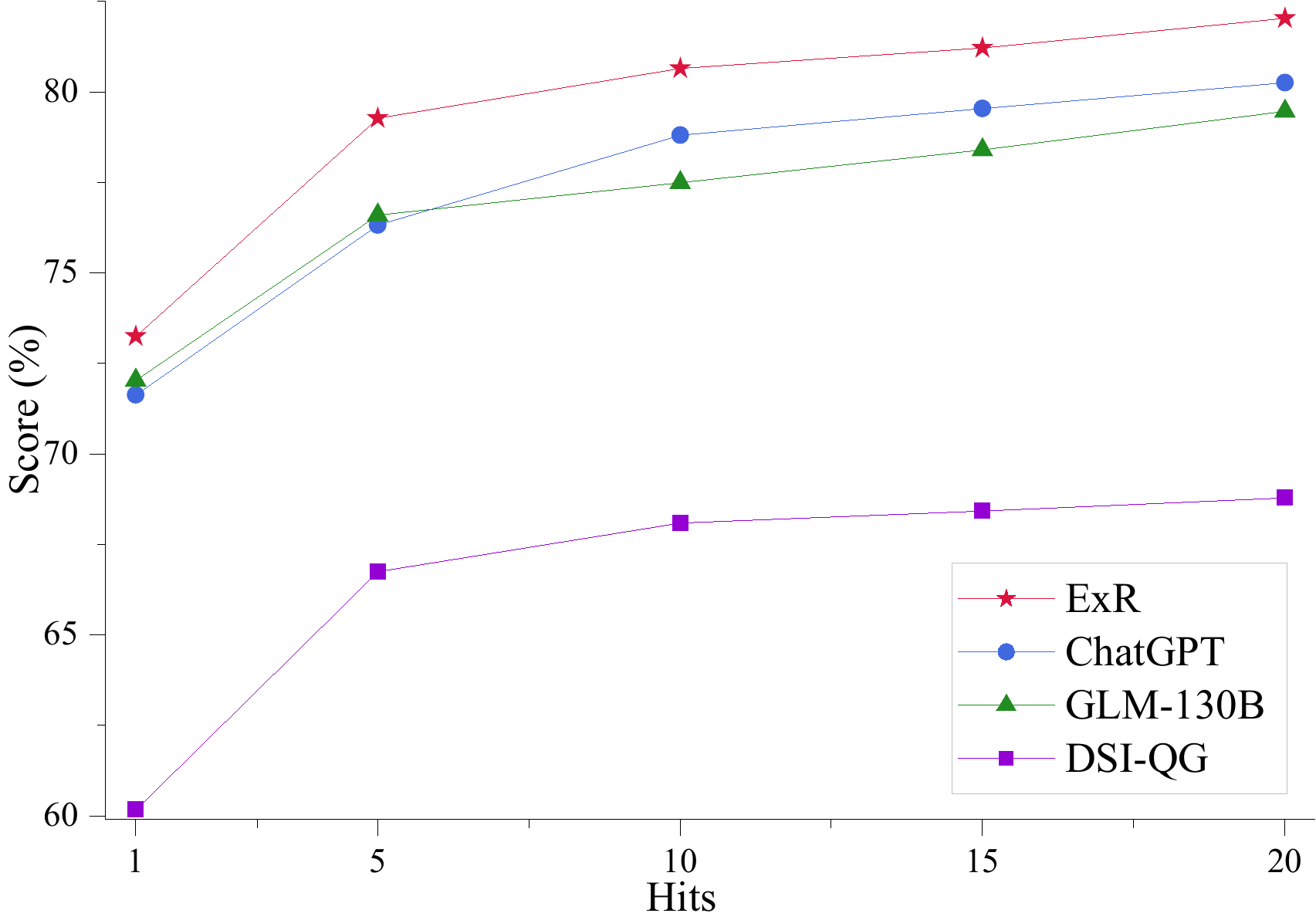}
   \caption{ Extraction strategies on NQ\_10K.}
   \label{exp:exr_nq}
\end{minipage}
\hfill
\begin{minipage}[t]{0.45\linewidth}
   \includegraphics[width = 1\linewidth]{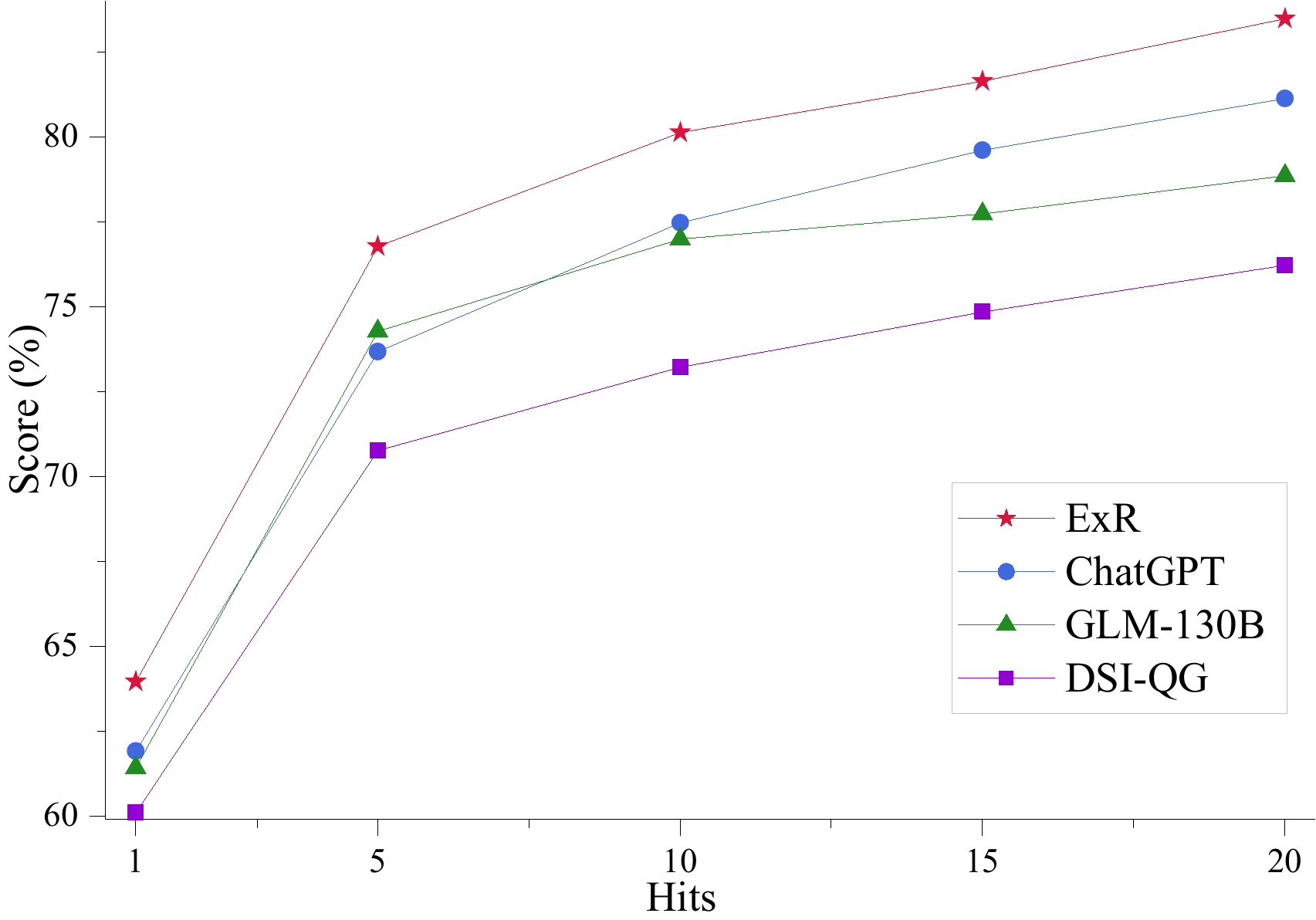}
   \caption{Extraction strategies on Dureader\_50K.}
   \label{exp:exr_dureader}
\end{minipage}
\end{tabular}
\end{figure}

\textbf{Does Event GDR have good generalization? } \  \textit{\textbf{Yes. Experimental results across various datasets, languages, and taxonomy settings demonstrate that our method exhibits good generalization and does not rely on any specific taxonomy. }} First, our method achieves significant results across datasets in different languages, including NQ\_10K in English and DuReader\_50K in Chinese. Second, we apply the taxonomy from MAVEN~\cite{MAVEN_wang_2020} to NQ\_10K and directly translate it to Chinese for DuReader\_50K without customizing it. The results show that our method also performs well on DuReader\_50K, indicating that it does not depend on a specific taxonomy. Besides, in following section, we also explore the impact of using various large models on event  extraction.

\subsection{Analysis on Knowledge Extraction Strategy}

To investigate the impact of various knowledge extraction strategies, we compare our Exchange-then-Reflection (ExR) with methods that solely rely on individual LLMs, such as \textit{ChatGPT} and \textit{GLM-130B}. From the results in Figure \ref{exp:exr_nq} and \ref{exp:exr_dureader}, we can see that: (1) our proposed ExR achieves superior performance compared to the variants using individual LLMs; (2) compared to DSI-QG, the model performance is improved whichever extraction strategy is used; 

\subsection{Scaling Model Size}

\begin{figure}[t]
\centering
\begin{tabular}{cc}
\begin{minipage}[t]{0.45\linewidth}
   \includegraphics[width = 1\linewidth]{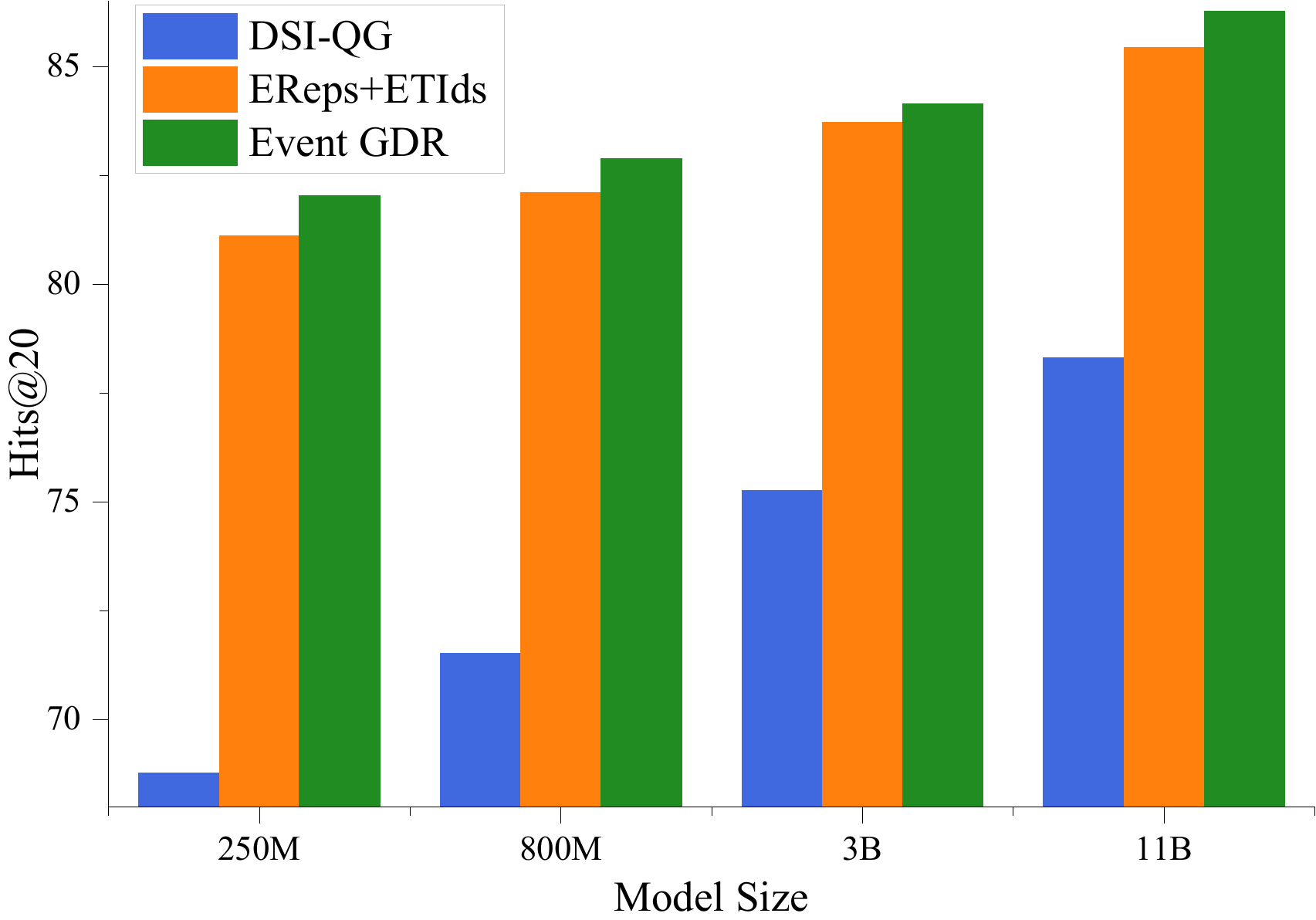}
   \caption{ Scaling model size on NQ\_10K.}
   \label{fig:model_size_nq}
\end{minipage}
\hfill
\begin{minipage}[t]{0.45\linewidth}
   \includegraphics[width = 1\linewidth]{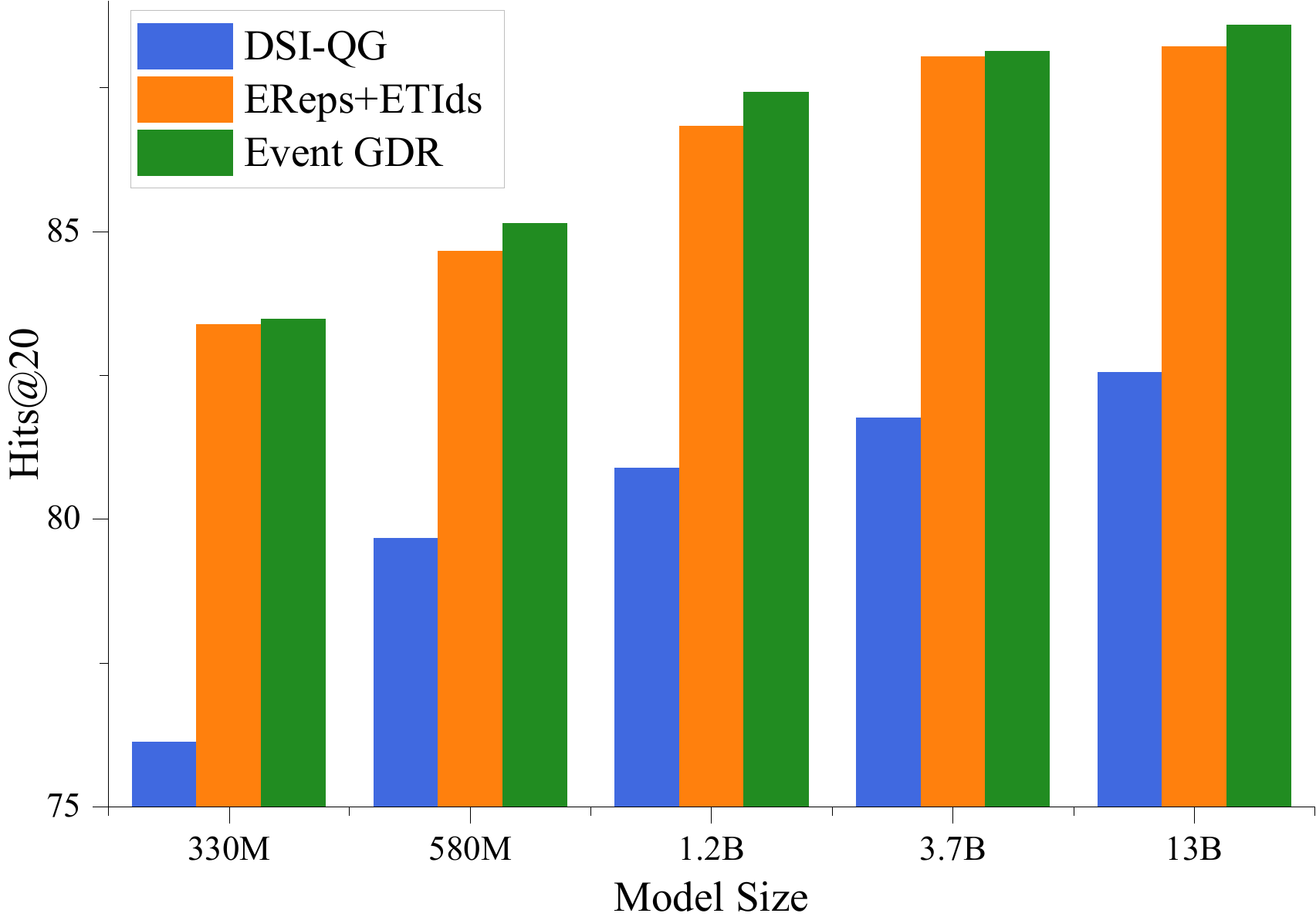}
   \caption{Scaling model size on Dureader\_50K.}
   \label{fig:model_size_dureader}
\end{minipage}
\end{tabular}
\end{figure}

To examine the effectiveness of Event GDR on different model sizes, we conduct extensive experiments on both datasets, and scale the model size up to 11B (T5-XXL) and 13B (mT5-XXL), respectively.
The results are shown in Figure \ref{fig:model_size_nq} and \ref{fig:model_size_dureader}, we can see that model performance continues to improve with increasing model size, and our approach consistently outperforms existing baselines.

\section{Conclusions}

In this paper, we propose Event GDR, an event-centric generative document retrieval model. We first utilize an exchange-then-reflection procedure based on multi-agents for event knowledge extraction. Then, we employ events and relations to model the document to guarantee the comprehensiveness and inner-content correlation. Finally, we map the events to well-defined taxonomy to construct the identifiers with explicit semantic structure. Experimental results on two widely used datasets indicate our proposed Event GDR substantially outperforms the strong baselines.

\begin{acks}
This work is supported by a grant from the Institute for Guo Qiang, Tsinghua University (2019GQB0003).
\end{acks}

\bibliographystyle{ACM-Reference-Format}
\balance
\bibliography{event-gdr}

\end{document}